\documentclass[journal, 10pt, twocolumn]{IEEEtran}

\usepackage{cite}
\usepackage{amsmath}
\usepackage{graphicx}
\usepackage{textcomp}
\usepackage{xcolor}
\usepackage{url}
\usepackage{siunitx}
\usepackage{booktabs}
\usepackage{multirow}
\usepackage{balance}

\begin{document}

\title{Protection Switching in Hybrid Hollow-Core and Single-Mode Fiber
Networks: Challenges, Analysis, and Mitigation Strategies}

\author{Md Ghulam Saber and Zhiping Jiang%
\thanks{M. G. Saber and Z. Jiang are with Ottawa Research Center, Huawei Technologies Canada, 303 Terry Fox Drive, Kanata, ON, K2K 3J1, Canada.
(e-mail: md.ghulam.saber@huawei.com).}}

\markboth{IEEE JOURNAL}{}

\maketitle

\begin{abstract}
Hollow-core fibers (HCF) are transitioning from laboratory curiosities to
production-deployed infrastructure, with major cloud providers operating
thousands of kilometers of hollow-core links.  As operators incrementally
upgrade their networks, working and protection paths will inevitably
traverse different fiber types, creating a new class of protection
switching challenges absent in homogeneous single-mode fiber networks.
This article provides a comprehensive overview of these
challenges and presents a comparative analysis of protection
switching under two architectures---1+1 dedicated and shared backup path
protection (SBPP)---in hybrid hollow-core and single-mode fiber
networks.  Using Monte Carlo simulation with random per-link fiber
assignment across six reference topologies
(1,602 node pairs), we quantify chromatic dispersion (CD) steps, generalized signal-to-noise ratio (GSNR) penalties, and modulation-format degradation for both architectures.
At 50\% HCF deployment, mean CD steps range
from 4,000 to 22,000~ps/nm, with GSNR penalties of 1.6--3.1~dB and
38--59\% of node pairs requiring modulation downgrade under 1+1
protection.  A complementary cross-fiber extreme analysis reveals
that the two switching directions are fundamentally asymmetric:
HCF$\rightarrow$SMF switching doubles the CD step and inflicts a
$\sim$10~dB GSNR penalty, while SMF$\rightarrow$HCF switching
delivers a \emph{negative} GSNR penalty (the protection path is
higher quality than the working path).  SBPP shows up to 7\% higher
CD steps and 4~percentage points (pp) more downgrade in sparsely connected
topologies due to its greedy shortest-first path selection.  Capacity
retention improves with HCF penetration for both architectures,
reaching 85--99\% at full HCF deployment.  We present mitigation strategies including DSP pre-loading,
spectral pre-equalization, and network planning guidelines, concluding
that 1+1 dedicated protection is preferable to SBPP for hybrid
deployments.
\end{abstract}

\section{Introduction}

Optical network protection switching has been a cornerstone of
telecommunications reliability for decades.  Under ITU-T G.808.1,
1+1 linear protection continuously bridges traffic onto both working
and protection paths, with a 50~ms switchover target.  In conventional
single-mode fiber (SMF) networks, this process is well understood:
both paths share the same fiber type, and the receiver's digital signal
processing (DSP) engine need only adapt to a different path length and
amplifier chain. The CD coefficient, nonlinear
behavior, and propagation characteristics exhibit negligible variation.

The emergence of hollow-core fiber (HCF) is fundamentally altering
this landscape.  Antiresonant nodeless fibers now achieve attenuation
below 0.1~dB/km~\cite{Petrovich2025}, while offering $\sim$30\% lower
latency, near-zero Kerr nonlinearity, and improved thermal
stability~\cite{Poggiolini2022_HCF}.  Commercial deployment is
accelerating: Microsoft operates over 1,280~km of HCF in production
Azure networks~\cite{microsoft2024}, and recent demonstrations include
full C-band 400G ZR bidirectional transmission over hollow-core
cable~\cite{hong_ofc2026}.

Crucially, research is now moving from physical-layer demonstrations
to network-level planning.  Ibrahimi~et~al.~\cite{ibrahimi_ofc2026}
showed that upgrading just 13.7\% of links to HCF reduces edge
data-center requirements by 24--50\%, while also enabling significant
savings in amplification power per Tbps through high-power EDFA
operation.  Pedro~et~al.~\cite{pedro_ofc2026} demonstrated that
allocating 10--20\% of spans to HCF in a pan-European topology
increases feasible 800G lightpaths by 36--100\%.
Zami~et~al.~\cite{zami_ofc2026} found that optimal EDFA output power
for HCF-based transparent networks is $\sim$26~dBm rather than the
maximum available, revealing non-trivial amplifier design trade-offs.
These studies collectively establish that \emph{hybrid} HCF--SMF
networks---where operators selectively upgrade links based on CAPEX
constraints---are the realistic near-term deployment scenario.

However, hybrid deployment introduces a fundamentally new class of
protection switching challenges.  HCF and SMF differ in virtually
every physical-layer parameter: the CD coefficient of HCF is
approximately 3.5~ps/(nm\,km) versus 17~ps/(nm\,km) for SMF.  HCF exhibits intermodal interference (IMI) from
residual higher-order modes at $-$40 to $-$56~dB/km~\cite{Fokoua2023},
absent in effectively single-mode SMF\@.  Residual CO$_2$ trapped in
the hollow core creates sharp narrowband absorption features
($\sim$1~GHz FWHM) that cause channel-specific OSNR penalties~\cite{he_ofc2026}.  HCF's negligible nonlinearity permits
per-channel launch powers of $\sim$+15~dBm (constrained by commercial EDFA output power, not by the fiber itself)---some 15~dB above SMF---creating
large GSNR asymmetries during switchover.

When protection switching occurs between paths of different fiber
types---SMF working to HCF protection, or vice versa---the receiver
DSP faces an abrupt, simultaneous change in accumulated CD, noise
profile, nonlinear characteristics, and latency.  Critically, these
two switching directions present \emph{asymmetric} challenges.
SMF-to-HCF switching involves lower accumulated CD on the protection
path but introduces IMI and CO$_2$ absorption; HCF-to-SMF switching
faces massive CD steps from the higher dispersion coefficient and
degraded GSNR due to the loss of HCF's launch-power advantage.
Neither direction has precedent in conventional network design.

This article provides the first comparative analysis of protection
switching architectures---1+1 dedicated and SBPP---in hybrid HCF--SMF
networks.  Our contributions include:
(1)~an overview of cross-fiber protection switching
challenges spanning CD steps, IMI transients, gas absorption, and
launch power asymmetry;
(2)~a Monte Carlo simulation framework with random per-link fiber
assignment (200 trials per scenario) across six reference topologies,
using per-link locally optimized launch
power~\cite{Poggiolini2022_HCF} and GN-model GSNR accumulation;
(3)~a direct comparison of 1+1 (Suurballe's joint-optimal
paths~\cite{Suurballe1984}) versus SBPP (greedy shortest-first path
selection), revealing that SBPP creates greater path asymmetry and
suffers trap-topology failures; and
(4)~mitigation strategies and network planning guidelines, including
a recommendation to prefer 1+1 over SBPP for hybrid deployments.

The remainder of this article is organized as follows.  We first
review optical network protection and the emerging HCF deployment
paradigm.  We then systematically catalog the challenges of
cross-fiber protection switching.  Our simulation methodology and
results follow, leading to mitigation strategies and a discussion of
open research challenges.

\section{Background: Protection Switching and the HCF Paradigm}

\subsection{Optical Network Protection Today}
\label{prot_mech}
Modern optical transport networks employ several protection
architectures, two of which dominate deployed mesh networks.

\textbf{1+1 dedicated protection} pre-provisions a protection path for
each working path, and the receiver continuously monitors both signals.
Path computation uses Suurballe's algorithm~\cite{Suurballe1984} with
Johnson's reweighting, which finds the \emph{minimum total cost} pair
of edge-disjoint paths.  This joint optimization may select a working
path that is slightly longer than the absolute shortest in order to
obtain a much better protection path, yielding balanced path lengths
and moderate DSP adaptation requirements at switchover.

\textbf{Shared Backup Path Protection (SBPP)} improves spectral
efficiency in homogeneous networks by allowing multiple working paths
to share backup capacity on links that are unlikely to fail
simultaneously.  The routing component of SBPP selects the working
path as the absolute shortest (Dijkstra), and the protection path as
the shortest link-disjoint alternative---a greedy two-step
computation.  Because the working path always claims the best links,
the protection path is forced to route around it, often traversing
significantly longer routes.  This greedy strategy can also fail on
\emph{trap topologies}, where a jointly optimal disjoint pair exists
but the shortest-first selection blocks it.  Furthermore, SBPP does
not pre-provision the protection path until a failure event triggers
activation; consequently, the receiver cannot pre-monitor the
protection signal, making DSP pre-loading more challenging.

We model the routing behavior of each
architecture---Suurballe's joint optimization for 1+1 and greedy
two-step Dijkstra for SBPP---but do not model backup capacity sharing,
which is outside the scope of this physical-layer study.  Each node
pair is analyzed independently, so the comparison isolates the effect
of the path selection algorithm on protection switching impairments.
The terms ``1+1'' and ``SBPP'' throughout this article refer
specifically to the routing strategies described above.

Both architectures use edge-disjoint paths (sharing no common links) to
survive single-link failures such as fiber cuts.  Upon detection of a
failure---typically via loss of signal or degraded bit error
ratio---the receiver switches to the protection path.  
%
\subsection{The Hollow-Core Fiber Revolution}

The nested antiresonant nodeless fiber (NANF) design, proposed by
Poletti~\cite{Poletti2014}, confines light within an air core
surrounded by a cladding of nested glass capillaries.  The
antiresonance mechanism suppresses leakage, while the nodeless
geometry eliminates surface mode coupling.  Subsequent refinements
--- including double-nested DNANF, support tube designs, and
fourfold-truncated structures --- have driven attenuation from
1.3~dB/km in 2019 down to 0.04~dB/km in 2026.

From a system perspective, HCF is characterized by several distinctive
properties~\cite{Fokoua2023,Poggiolini2022_HCF}:

\begin{itemize}
\item \textbf{Low CD:} 2--4~ps/(nm\,km), approximately $5\times$
  lower than SMF.
\item \textbf{Near-zero nonlinearity:} Kerr coefficient
  $\gamma \approx 0.001$~W$^{-1}$km$^{-1}$, versus 1.3 for SMF.
\item \textbf{Low latency:} Effective index $n_\text{eff} \approx 1.0003$
  versus 1.468 for SMF.
\item \textbf{IMI:} Residual higher-order modes create intermodal
  interference at levels of $-$30 to $-$56~dB/km. Lowest reported to date $-$73.3~dB/km.
\item \textbf{Gas absorption:} Trapped CO$_2$, CO, and H$_2$O produce
  narrowband absorption peaks across the telecom bands.
\item \textbf{Ultra-low Rayleigh backscatter:} Up to 30~dB lower than
  SMF, enabling same-wavelength bidirectional transmission.
\end{itemize}

Recent network planning studies~\cite{ibrahimi_ofc2026,pedro_ofc2026} have demonstrated the strategic
value of selective HCF deployment.  

\subsection{Hybrid HCF--SMF: The Transition Reality}

Complete overnight replacement of SMF with HCF is neither economically
feasible nor technically necessary.  The transition will be gradual,
creating hybrid networks where HCF and SMF coexist on different links.
This heterogeneity is actually desirable from a cost perspective:
deploying HCF preferentially on high-traffic or latency-sensitive links
captures most of the performance benefit while minimizing capital
expenditure.

However, this hybrid paradigm creates an asymmetry that is invisible
during normal operation but becomes critical during protection events:
the working and protection paths may use entirely different fiber types.
This article focuses specifically on this protection switching asymmetry
and its consequences for the two most common protection architectures.

\section{Challenges of Cross-Fiber Protection Switching}

When a protection switch occurs between paths of different fiber types,
the receiver DSP confronts simultaneous discontinuities in multiple
physical-layer parameters.  We categorize these challenges into five
areas.

\subsection{Chromatic Dispersion Step}

A key challenge is the abrupt CD change. Consider an SMF
working path of length $L_w$ switching to an HCF protection path of
length $L_p$.  The accumulated CD changes from $17 \times L_w$ to
$3.5 \times L_p$~ps/nm.  For typical backbone paths of 500--3,000~km,
this creates CD steps of 4,000 to over 22,000~ps/nm --- far exceeding
the tracking range of blind CD equalizers, which are designed for
gradual drift compensation, not instantaneous jumps of this magnitude.

The number of frequency-domain equalizer (FDE) taps required to
compensate accumulated CD scales linearly with the CD
magnitude~\cite{Savory2010}:
\begin{equation}
N_\text{taps} = \left\lceil \frac{|\text{CD}| \cdot \Delta\lambda_\text{sig}}{T/2} \right\rceil + 1
\label{eq:fde}
\end{equation}
where $\Delta\lambda_\text{sig}=\lambda^2 R_s/c$ is the signal spectral
width and $R_s$ is the symbol rate.  At 64~GBaud with $2\times$ oversampling, every 1,000~ps/nm
requires approximately 67 additional $T/2$-spaced taps.  A CD step of
20,000~ps/nm thus requires the equalizer to re-acquire approximately
1,300 additional taps---a substantial increase in the adaptation range
that challenges blind CD estimation algorithms.

In a same-fiber switch (SMF-to-SMF or HCF-to-HCF), the CD step is
determined solely by path length difference, which is typically
modest.  The cross-fiber case is qualitatively different because
the dispersion coefficient itself changes by a factor of five.

\subsection{Intermodal Interference Transient}

When switching from an SMF path (zero IMI) to an HCF path, the DSP encounters
a qualitatively different impairment: energy from higher-order modes leaks
coherently into the detected fundamental mode at delays set by the differential
group delay between modes.
Unlike polarization crosstalk---which a $2\times2$ MIMO equalizer can invert
because both polarizations are received---the higher-order modes are not
individually detected at the coherent receiver.
As a result, IMI cannot be cancelled by standard DSP; it manifests as a
stochastic, signal-correlated noise floor that accumulates with distance and is
not recoverable through equalization.
The only effect the adaptive equalizer can have is to partially suppress the
deterministic inter-symbol interference (ISI) component of IMI by extending its
tap memory to span the inter-modal differential
group delay, but this increases the equalizer
adaptation burden without removing the fundamental GSNR penalty. Research demonstrations with mode-selective front ends (e.g., photonic-lantern-based multi-mode coherent
receivers) can in principle cancel IMI through per-mode MIMO; however, this approach is practically prohibitive for standard telecom deployments.

\subsection{Gas Line Absorption: The CO$_2$ Challenge}
\label{co2_discuss}
A unique impairment in HCF is the absorption by residual gas molecules
trapped within the hollow core.  CO$_2$ is the most problematic
species, producing multiple sharp absorption lines across the
L-band~\cite{chen_ecoc2025}.  These lines have Lorentzian profiles
with full-width at half-maximum of approximately 1~GHz and spacing
of approximately 30--50~GHz.  Reported peak absorption values span
a wide range, from $\sim$0.08 to $\sim$0.2~dB/km depending on HCF
design, gas partial pressure, and measurement
bandwidth~\cite{chen_ecoc2025,he_ofc2026}.  One-shot field
measurement techniques can characterize the resulting per-channel
OSNR penalty~\cite{he_ofc2026}, enabling pre-computation of
mitigation parameters.

For protection switching, gas absorption creates two distinct problems.
First, the protection path's absorption spectrum may differ from the
working path's spectrum (or be entirely absent if the working path is
SMF), requiring the receiver DSP to adapt to a new spectral
environment.  Second, certain WDM channels may experience
significantly higher penalties than others on the protection path,
potentially causing channel-dependent outages that are difficult to
predict without detailed spectral characterization.

Recent work has demonstrated DSP-based mitigation techniques.
However, these mitigation techniques assume knowledge of
the absorption profile.  During a protection switch to an HCF path,
the CO$_2$ characteristics must either be pre-characterized and
stored, or rapidly estimated by the DSP in real time which is an open research area.

\subsection{Launch Power and Amplifier Asymmetry}

The optimal per-channel launch power differs significantly between
fiber types.  For SMF, the locally optimized (LOGO) launch
power~\cite{Poggiolini2022_HCF} balances amplified spontaneous
emission (ASE) noise against Kerr nonlinear interference (NLI),
typically yielding values around 0 to +1~dBm per channel.  For HCF,
the near-zero nonlinearity means that higher launch powers are always
beneficial up to the practical EDFA limit of approximately +15~dBm per
channel (approximately +33~dBm total for 64
channels)~\cite{Poggiolini2022_HCF}.

Upon switchover, the EDFA chain on the protection path operates at
different gain and output power levels.  If the protection path
uses HCF, its amplifiers may be configured for higher output powers,
and the gain transient during sudden traffic loading can cause
temporary power excursions that further perturb the DSP.
Zami~et~al.~\cite{zami_ofc2026} showed that exceeding 30~dBm EDFA
output power in HCF networks actually worsens power efficiency,
suggesting that practical HCF deployments will use amplifier
configurations distinct from SMF networks.

\begin{figure*}[!t]
	\centering
	\includegraphics[width=0.8\textwidth]{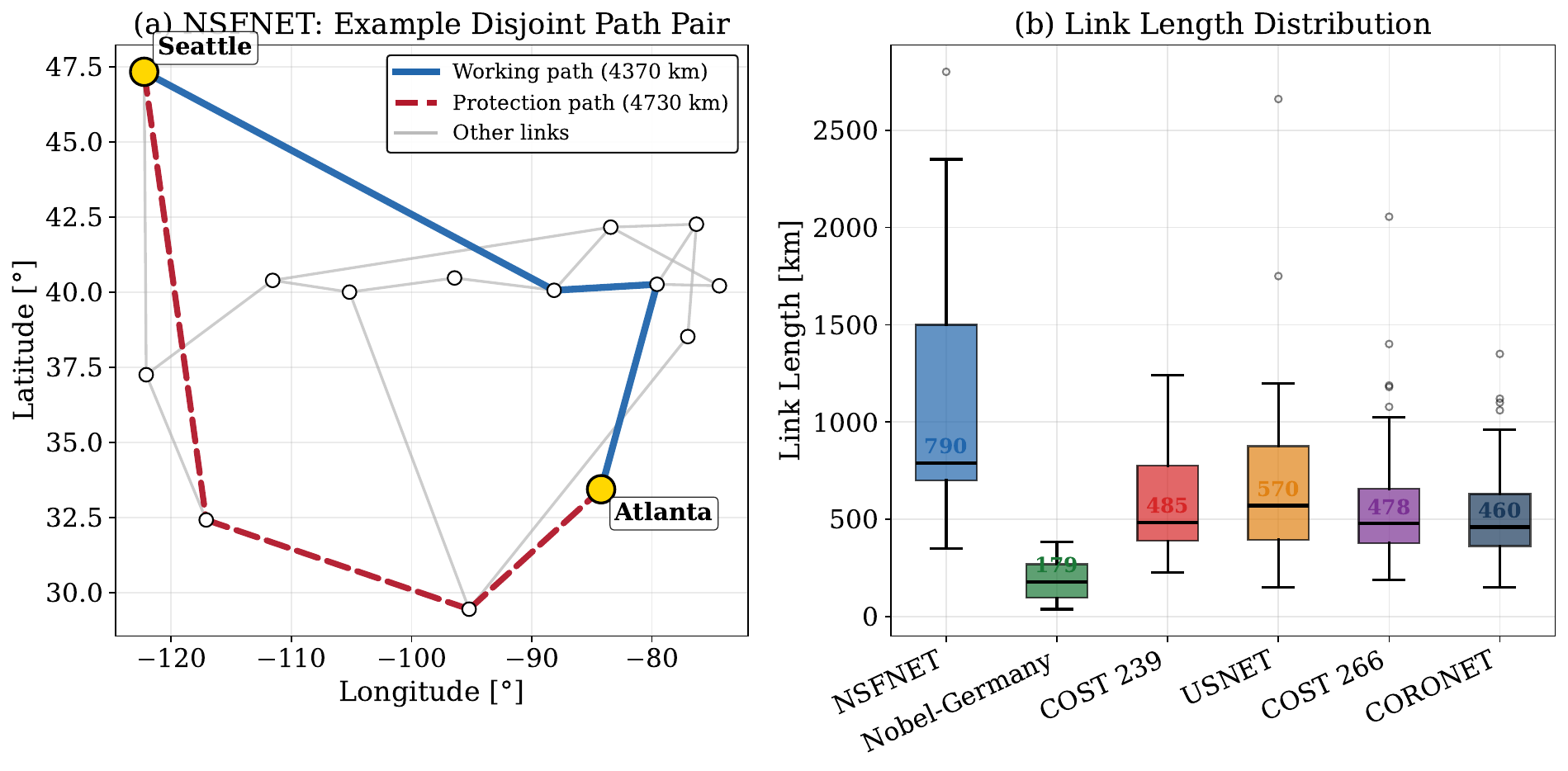}
	\caption{(a)~Example link-disjoint path pair for the
		Seattle--Atlanta node pair in NSFNET, computed by Suurballe's
		algorithm: blue solid line is the working path (4,370~km), red dashed
		line is the edge-disjoint protection path (4,730~km), gray lines are
		unused links.  (b)~Link length distribution per topology
		(box: interquartile range; whiskers: 1.5$\times$Interquartile Range; circles:
		outliers; median annotated). }
	\label{fig:links}
\end{figure*}

\subsection{Propagation Delay Asymmetry}
\label{sec:delay_asymmetry}
A challenge unique to hybrid HCF--SMF networks is the propagation
delay mismatch between fiber types. SMF has a group delay of
$\sim$4.9~$\mu$s/km while HCF propagates at $\sim$3.3~$\mu$s/km,
a difference of $\sim$1.6~$\mu$s/km. In conventional all-SMF
networks, the differential delay between working and protection paths
is determined solely by path length asymmetry. In hybrid networks,
the fiber-type composition introduces a second degree of freedom:
HCF on a longer protection path partially compensates for the greater
distance through faster propagation, while HCF on a shorter working
path paired with an SMF protection path produces the worst-case
differential delay. This has direct consequences for hitless
protection switching, which requires buffering data on the
shorter-delay path until the longer-delay path's copy arrives ---
buffer depth scales linearly with differential delay, making
fiber-type composition a non-trivial factor in transceiver buffer
dimensioning.

Furthermore, because links are upgraded from SMF to HCF
incrementally, the differential delay profile of both working and
protection paths evolves stochastically over the network lifetime.
Protection equipment must therefore support a wide range of
differential delays, or the network management system must
dynamically update buffer alignment parameters as the fiber inventory
evolves.
\begin{table}[!t]
	\centering
	\caption{Structural Properties of Reference Topologies}
	\label{tab:topo}
	\setlength{\tabcolsep}{2.5pt}
	\begin{tabular}{lcccccc}
		\toprule
		\textbf{Topology} & $N$ & $E$ & \textbf{Pairs} & $\bar{d}$
		& $\bar{L}_W$ & $\bar{L}_P/\bar{L}_W$ \\
		\midrule
		NSFNET      & 14 & 21 &   91 & 3.0 & 2,249 & 1.65 \\
		Nobel-Ger.  & 17 & 26 &  136 & 3.1 &   447 & 1.69 \\
		COST~239    & 11 & 26 &   55 & 4.7 &   928 & 1.27 \\
		USNET       & 24 & 43 &  276 & 3.6 & 2,087 & 1.29 \\
		COST~266    & 37 & 57 &  666 & 3.1 & 1,953 & 1.51 \\
		CORONET     & 30 & 48 &  378 & 3.2 & 2,044 & 1.47 \\
		\bottomrule
	\end{tabular}
	\vspace*{0.15cm}
	
	{\footnotesize $N$: nodes, $E$: links,
		$\bar{d} = 2E/N$: mean node degree (average number of links per node),
		$\bar{L}_W$: mean working-path length~[km],
		$\bar{L}_P/\bar{L}_W$: mean protection-to-working path length ratio (1+1~protection).}
\end{table}
%
%

\section{Multi-Topology Simulation Study}
\begin{figure*}[!t]
	\centering
	\includegraphics[width=\textwidth]{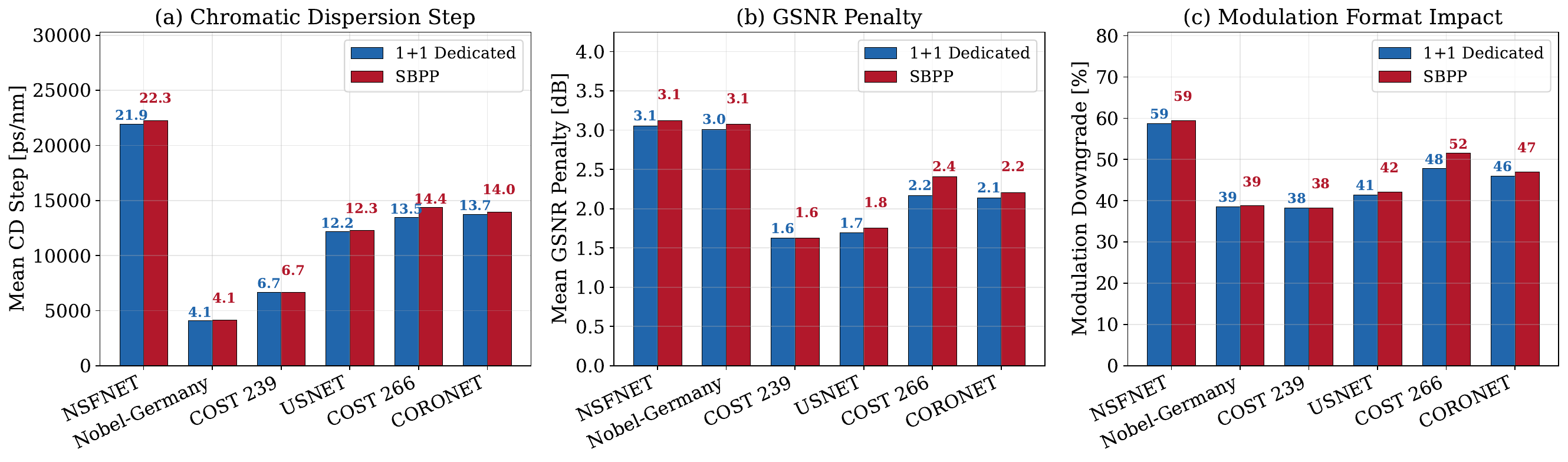}
	\caption{Architecture comparison at 50\% HCF (200 MC trials):
		(a)~mean CD step, (b)~mean GSNR penalty, and (c)~modulation
		downgrade for 1+1 dedicated protection (blue) vs.\ SBPP (red).
		SBPP shows moderately higher impairments in sparsely connected
		topologies; dense topologies (COST~239) yield identical results.}
	\label{fig:arch}
\end{figure*}
\subsection{Methodology}

To quantify the protection switching challenges across realistic
network geometries, we developed a Monte Carlo simulation platform that
models the complete physical-layer impairment chain for hybrid HCF--SMF
networks under both 1+1 and SBPP protection architectures.

\textbf{Network topologies:} We evaluate six standard reference
topologies of increasing size and connectivity: NSFNET (14 nodes,
21 links), Nobel-Germany (17 nodes, 26 links), COST~239 (11 nodes,
26 links), USNET (24 nodes, 43 links), COST~266 (37 nodes, 57 links),
and CORONET (30 nodes, 48 links).  Table~\ref{tab:topo} summarizes
their structural properties; Fig.~\ref{fig:links} shows an example
disjoint path pair (NSFNET Seattle--Atlanta) computed by Suurballe's
algorithm and the link length distributions for all six topologies.

\textbf{Path computation:} For 1+1 dedicated protection, we compute
the minimum-cost pair of edge-disjoint paths using Suurballe's
algorithm~\cite{Suurballe1984} with Johnson's reweighting, which
jointly optimizes both paths.  For SBPP, we use the greedy two-step
Dijkstra routing described in Section~\ref{prot_mech}: the working path is the
absolute shortest, and the protection path is the shortest
link-disjoint alternative.

Under 1+1, all six topologies yield 1,602 node pairs with valid
disjoint paths.  Under SBPP, COST~266 loses 2 pairs
(Copenhagen--Krakow and Krakow--Oslo) due to the trap-topology effect.
This is not a topological connectivity failure---a pair of
link-disjoint paths exists, as found by Suurballe's algorithm---but a
routing artifact: the greedy two-step approach cannot discover the
disjoint pair because the shortest-first working path consumes links
that would be needed for the backup route.

\textbf{Monte Carlo simulation:} In real hybrid deployments, both
working and protection paths traverse a mix of SMF and HCF links
rather than consisting entirely of one fiber type.  To capture this
heterogeneity, we employ Monte Carlo simulation: at each trial, every
link is independently assigned as HCF with probability equal to the
target deployment fraction, or SMF otherwise.  Each topology is
constructed deterministically from its standard reference definition;
the random seed controls only the per-link fiber-type assignment.
Both protection architectures are evaluated on the same random
realization per trial, ensuring a fair comparison.  We run 200 trials
per topology per deployment fraction per architecture.  A convergence
analysis confirmed that 200 trials yields a standard error of the mean
below 1.5\% for all key metrics.

\textbf{Physical-layer model:} Link distances are taken from standard
reference topology definitions.  Each link is subdivided into uniform
amplifier spans of 80~km (SMF) or 100~km (HCF), with inline EDFAs
(NF~$=$~5.5~dB) and 64 WDM channels at 75~GHz spacing with
64~GBaud symbol rate. The fiber parameters adopted throughout the
study are representative of commercially available fibers: for SMF,
attenuation 0.20~dB/km, CD coefficient 17~ps/(nm$\cdot$km),
nonlinear coefficient $\gamma=1.3$~W$^{-1}$km$^{-1}$, effective index
$n_\text{eff}=1.468$, and PMD coefficient
0.06~ps/$\sqrt{\text{km}}$; for HCF, attenuation 0.13~dB/km, CD
coefficient 3.5~ps/(nm$\cdot$km), $\gamma=0.001$~W$^{-1}$km$^{-1}$,
$n_\text{eff}=1.0003$, PMD coefficient
0.05~ps/$\sqrt{\text{km}}$, and IMI level $-$55~dB/km.
The GSNR per link is computed using the Gaussian noise
model for NLI~\cite{Poggiolini2022_HCF}, with per-link locally
optimized per-channel launch power (LOGO): approximately 0~dBm for SMF
and +15~dBm for HCF\@.  For heterogeneous paths containing both fiber
types, the end-to-end GSNR is accumulated segment-by-segment:
$\text{GSNR}_\text{tot}^{-1}=\sum_k \text{GSNR}_k^{-1}$.

\textbf{GSNR penalty:} We define the GSNR penalty as
$\Delta\text{GSNR} = \text{GSNR}_\text{work} - \text{GSNR}_\text{prot}$
in dB.  A positive penalty means the protection path has worse signal
quality than the working path.

\textbf{Modulation format:} The achievable format is determined by
comparing GSNR against standard thresholds: DP-64QAM ($\geq$24~dB),
DP-16QAM ($\geq$18~dB), DP-8QAM ($\geq$14~dB), DP-QPSK ($\geq$11~dB).

\subsection{Architecture Comparison at 50\% HCF Deployment}

Figure~\ref{fig:arch} compares the mean CD step, GSNR penalty, and
modulation downgrade for 1+1 and SBPP at 50\% HCF deployment.

\textbf{CD Step (Fig.~\ref{fig:arch}(a)):}
For most topologies, SBPP produces moderately higher mean CD steps
than 1+1.  The largest relative difference appears in COST~266
(14,374 vs.\ 13,451~ps/nm, +7\%), where SBPP's greedy
shortest-first routing creates less balanced path pairs.
NSFNET, USNET, and CORONET show 1--2\% differences.
Notably, COST~239 yields \emph{identical} results under both
architectures (6,678~ps/nm), because its dense connectivity
(degree~4.7) means Suurballe's algorithm and two-step Dijkstra
find the same disjoint path pairs.

\textbf{GSNR Penalty (Fig.~\ref{fig:arch}(b)):}
Both architectures show positive mean GSNR penalties (1.6--3.1~dB),
confirming that working paths on average achieve higher GSNR than
protection paths.  The penalty is positive because the working
path---always shorter---accumulates less noise, and in mixed-fiber
paths may benefit from more HCF links with their higher LOGO launch
power.  SBPP penalties are marginally higher (by 0.1--0.2~dB) in
sparsely connected topologies, while dense topologies show negligible
differences.
\begin{figure*}[!t]
	\centering
	\includegraphics[width=0.9\textwidth]{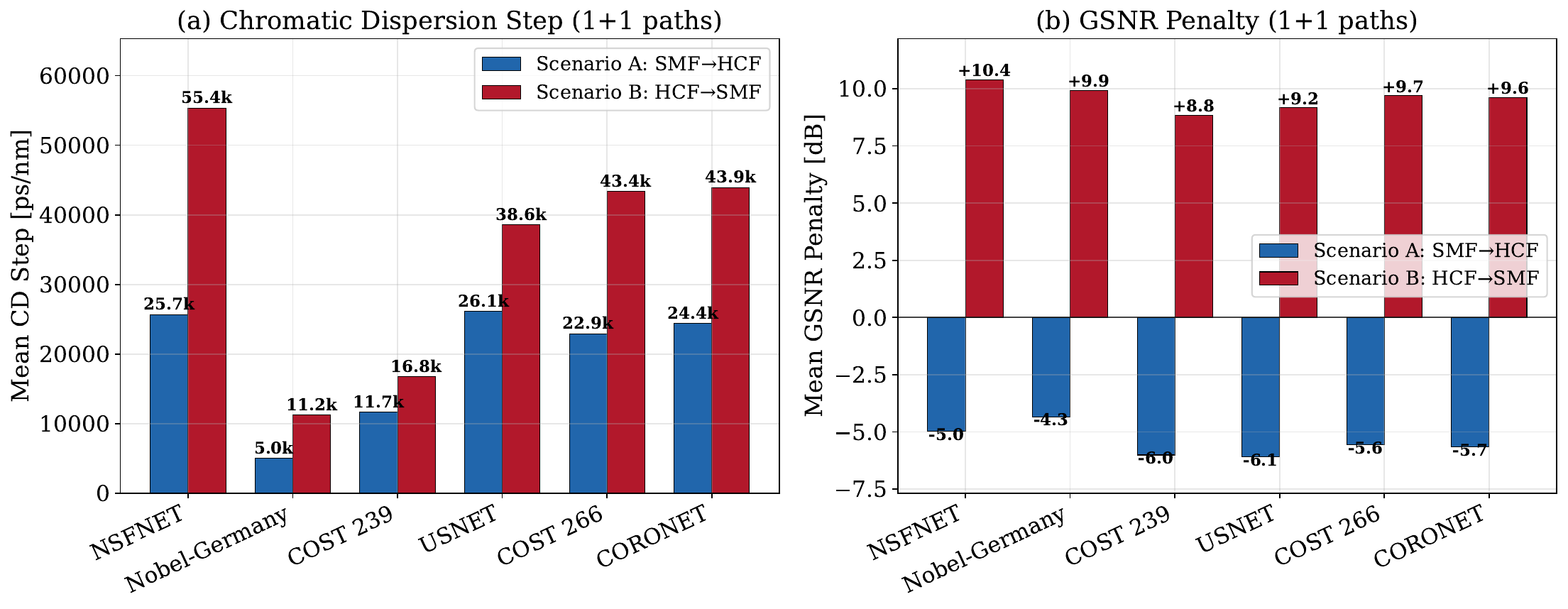}
	\caption{Cross-fiber extreme switching bounds under 1+1 dedicated
		protection: (a)~mean CD step and (b)~mean GSNR penalty for
		Scenario~A (all-SMF working $\rightarrow$ all-HCF protection, blue)
		and Scenario~B (all-HCF working $\rightarrow$ all-SMF protection,
		red).  Scenario~A delivers a negative GSNR penalty (the protection
		path is higher quality than the working path); Scenario~B doubles the
		CD step and inflicts a $\sim$10~dB GSNR penalty.}
	\label{fig:cfe}
\end{figure*}

\textbf{Modulation Downgrade (Fig.~\ref{fig:arch}(c)):}
At 50\% HCF, 1+1 produces 38--59\% modulation downgrade across
topologies.  SBPP shows up to 4 pp more downgrade
(e.g., COST~266: 52\% vs.\ 48\%), with the gap most pronounced in
topologies having lower connectivity.  Dense topologies (COST~239)
show no difference.  The downgrade occurs because the protection-path
GSNR is insufficient to support the same modulation format as the
working path.  NSFNET shows the highest downgrade rates under both
architectures (59\% for 1+1) because its sparse topology (average
degree 3.0) produces the most extreme path-length asymmetries, with
protection paths often 3--10$\times$ longer than working paths.

\textbf{Effect of HCF deployment fraction:}
The 50\% case above represents the intermediate regime; the deployment
endpoints reveal how the underlying CD coefficient drives impairment.
At 0\% HCF (legacy all-SMF baseline), CD steps and GSNR penalties stem
purely from path-length asymmetry: under 1+1, mean CD steps range from
4,200~ps/nm (COST~239) to 25,000~ps/nm (NSFNET), with 29--63\% of node
pairs requiring modulation downgrade.  At 100\% HCF (full deployment),
HCF's $5\times$ lower CD coefficient compresses these CD steps by
roughly the same factor, to 870--5,150~ps/nm, and modulation downgrade
falls sharply to 4--40\%.  The most asymmetric topology (NSFNET)
remains the worst across the entire HCF sweep, but its mean CD step
drops from 25,000 to 5,150~ps/nm and its downgrade rate from
63\% to 40\%---a substantial improvement that motivates full HCF
deployment as the long-term target rather than indefinite hybrid
operation.  SBPP tracks 1+1 closely at both endpoints (within
1--3\, pp of modulation downgrade), confirming that the architecture
gap arises predominantly at intermediate fractions where fiber-type
randomness amplifies the routing asymmetry.

\subsection{Cross-Fiber Extreme Switching Bounds}
\label{sec:cross_fiber}
The MC results above sample many possible per-link fiber assignments
at a fixed HCF fraction, but the worst-case impairments occur when
the working and protection paths consist \emph{entirely} of different
fiber types.  Figure~\ref{fig:cfe} bounds these extremes for 1+1
dedicated protection by computing, for every node pair in each
topology, the deterministic CD step and GSNR penalty under two
scenarios: \textbf{Scenario~A}, an all-SMF working path switching to
an all-HCF protection path (forward direction), and
\textbf{Scenario~B}, the reverse direction (all-HCF working to
all-SMF protection).  Together they bracket the impairment envelope
of any random per-link MC realization at intermediate HCF fractions.

\textbf{Scenario A (SMF$\rightarrow$HCF):} The protection path
benefits from HCF's $5\times$ lower CD coefficient and its $\sim$15~dB
higher LOGO launch power.  Mean CD steps remain large in absolute
terms (5,000--26,100~ps/nm across topologies, since the geometric
asymmetry of the path pair still applies), but the GSNR penalty
becomes \emph{negative} ($-4.3$ to $-6.1$~dB)---the protection
path is in fact \emph{higher} quality than the working path.
SMF$\rightarrow$HCF switching is therefore an opportunity rather
than a threat: the receiver must still re-acquire the equalizer for
the new CD value, but it can subsequently support an
\emph{upgraded} modulation format.

\textbf{Scenario B (HCF$\rightarrow$SMF):} This is the worst case for
both metrics.  Mean CD steps inflate to 11,200--55,400~ps/nm
(approximately $2\times$ Scenario~A), because the protection path on
SMF accumulates dispersion at the higher 17~ps/(nm$\,$km) coefficient
while the working path on HCF accumulated little to begin with.
GSNR penalties become large and positive ($+8.8$ to $+10.4$~dB),
reflecting the loss of HCF's launch-power advantage on the protection
path.  Both effects compound: the equalizer must adapt to a much
larger CD range \emph{and} the protection signal arrives substantially
weaker.  NSFNET, with a 1.65$\times$ mean protection-to-working
length ratio, exhibits the largest CD step (mean 55,400~ps/nm
$\approx$3,700 additional FDE taps via Eq.~\eqref{eq:fde}), with
worst individual pairs exceeding 80,000~ps/nm.

\begin{figure*}[!t]
	\centering
	\includegraphics[width=0.85\textwidth]{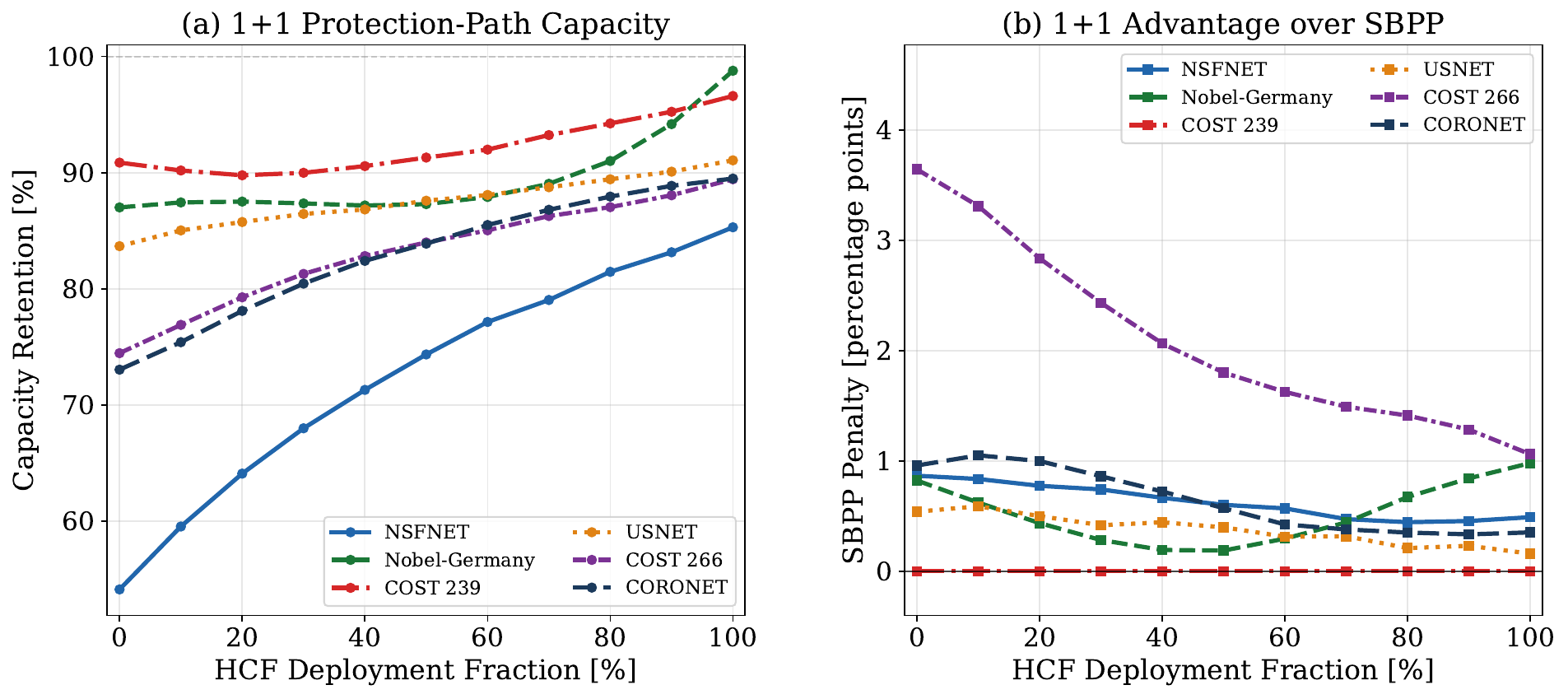}
	\caption{Two-panel view of capacity retention (200 MC trials per
		point).  (a)~1+1 dedicated protection: capacity retention vs.\ HCF
		deployment fraction for all six topologies, showing monotonic
		improvement from 54--91\% at 0\% HCF to 85--99\% at 100\% HCF.
		(b)~SBPP penalty: the gap between 1+1 and SBPP retention in
		pp, peaking at intermediate HCF fractions where
		fiber-type randomness amplifies the routing asymmetry.  COST~266
		exhibits the largest gap ($\sim$2--3 points), confirming 1+1's
		architectural advantage in sparsely connected topologies.}
	\label{fig:cap}
\end{figure*}

This directional asymmetry is the central qualitative result of the
cross-fiber switching analysis: the two switching directions are not
mirror images of one another but present fundamentally different
challenges and opportunities.  Random per-link MC results
(Fig.~\ref{fig:arch}) sit between Scenarios~A and~B at intermediate
HCF fractions; the asymmetry between them explains why mean CD
steps and GSNR penalties scale almost linearly with HCF fraction in
Fig.~\ref{fig:cap}, with sensitivity governed by the deployment
direction.

\subsection{Network Capacity Impact}

Figure~\ref{fig:cap} quantifies the aggregate capacity retention
after protection switching---defined as the ratio of total
protection-path throughput to working-path throughput---as a function
of HCF deployment fraction.  The throughput per node pair is determined
by the highest achievable modulation format at the protection-path
GSNR: DP-64QAM (600~Gb/s), DP-16QAM (400~Gb/s), DP-8QAM (300~Gb/s),
DP-QPSK (200~Gb/s).  This metric captures the modulation-format
downgrade penalty as a fractional throughput loss; it does not model
spectral efficiency or wavelength-level capacity planning.

Both architectures show improving capacity retention with increasing
HCF fraction, as HCF's lower loss and near-zero nonlinearity boost
protection-path GSNR.  At 0\% HCF (all-SMF baseline), retention
ranges from $\sim$55\% (NSFNET) to $\sim$91\% (COST~239), reflecting
the inherent path-length asymmetry of each topology.  Dense topologies
(COST~239, Nobel-Germany) achieve higher baseline retention because
their low $\bar{L}_P/\bar{L}_W$ ratios mean the protection path is
only modestly longer than the working path; the protection-path
GSNR therefore remains close enough to the working-path GSNR that
most node pairs retain their original modulation format.  Continental
topologies with extreme outliers (NSFNET) lose the most capacity
because their longest protection paths fall multiple thresholds below
the working-path modulation.

At 100\% HCF, retention reaches 85--99\% across all topologies, with
COST~239 and Nobel-Germany approaching near-full retention
($\sim$96--99\%).  NSFNET exhibits the largest absolute improvement
(55\% $\rightarrow$85\%, +30~pp) but remains the lowest in absolute
terms due to its extreme path-length outliers.  The improvement is
largely monotonic with HCF fraction for all topologies; small
non-monotonicities (e.g., COST~239 between 0\% and 30\%) reflect
trial-to-trial variance from the random per-link assignment.
Importantly, the curves do not saturate at any intermediate fraction:
each additional 10~pp of HCF deployment yields a measurable retention
gain, supporting the case for staged rollouts even when full
deployment is not immediately feasible.


The 1+1 vs.\ SBPP gap is modest: typically below 1 pp,
with COST~266 showing the largest gap ($\sim$2 points at 50\% HCF,
up to $\sim$3 points at 20--30\% HCF).  This occurs because SBPP,
with its 666 demand pairs, must route many protection paths away from
their individually shortest disjoint route to exploit backup sharing,
incurring GSNR penalties that manifest as modulation-format downgrades
and reduce aggregate capacity retention relative to~1+1.
COST~239 shows no difference, as its small demand set (55 pairs) and
dense connectivity leave SBPP little room to deviate from the same
paths selected by~1+1.  

\subsection{L-band CO$_2$ Channel Impact}

While the preceding analysis assumed C-band operation, many operators
are planning L-band expansion to alleviate C-band exhaustion.  The
CO$_2$ absorption lines discussed in Section~\ref{co2_discuss} lie predominantly
in the L-band~\cite{chen_ecoc2025,he_ofc2026}, so HCF protection paths
are far more sensitive to gas absorption in the L-band than in the
C-band.  We therefore quantify the channel-level impact of CO$_2$
absorption on the protection-path L-band GSNR at 50\% HCF deployment
under both architectures.

For each MC trial, we compute the per-channel CO$_2$ penalty for the
protection-path HCF length (linearly scaled from a reference profile
calibrated to~\cite{chen_ecoc2025}), subtract it from the CO$_2$-free
L-band GSNR, and count the number of channels that fall below a lower
modulation-format threshold than the CO$_2$-free baseline.  A channel
is thus classified as ``CO$_2$-affected'' only when its absorption
penalty is large enough to force a modulation downgrade on that
specific wavelength.

\begin{figure}[!t]
	\centering
	\includegraphics[width=\columnwidth]{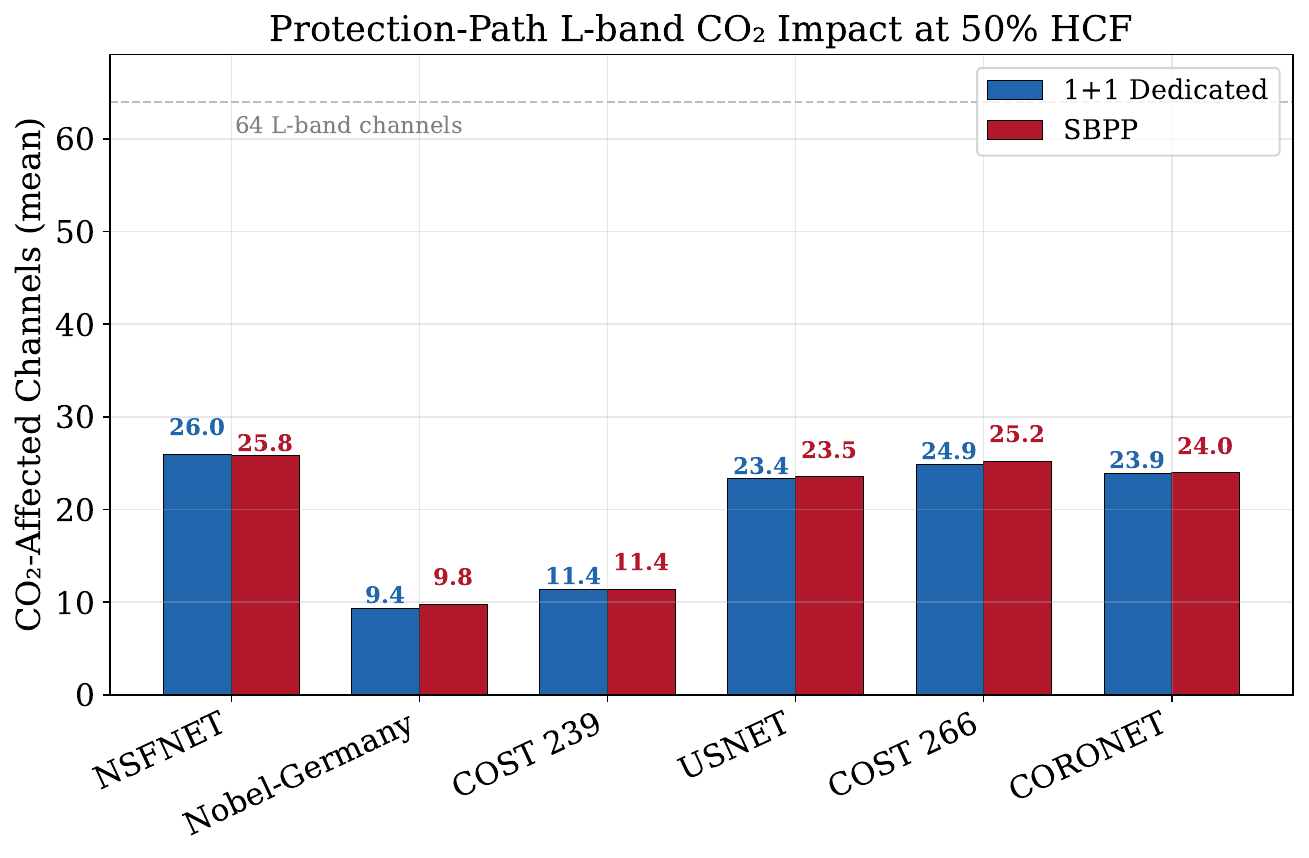}
	\caption{Mean number of L-band channels on the protection path whose
		achievable modulation format is downgraded by CO$_2$ absorption, at
		50\% HCF (200 MC trials).  Out of 64 L-band channels, continental
		topologies lose 24--26 channels on average; dense regional topologies
		lose only 9--12.}
	\label{fig:lband}
\end{figure}

Figure~\ref{fig:lband} shows the mean number of CO$_2$-affected channels
per topology out of 64 L-band channels.  Continental backbones with
long HCF path segments (NSFNET, USNET, COST~266, CORONET) see
24--26 channels affected on average, corresponding to $\sim$38--41\%
of the L-band grid.  Dense regional topologies with shorter HCF routes
(Nobel-Germany, COST~239) remain below 12 channels ($\sim$15--18\%).
SBPP and 1+1 produce nearly identical CO$_2$ impact---the differences
are below 0.4 channels in all topologies---because the L-band CO$_2$
penalty is dominated by the \emph{length} of HCF on the protection
path, not by the routing algorithm. Both architectures exhibit similar mean HCF lengths on protection
paths at a fixed deployment fraction, confirming that the CO$_2$
penalty is architecture-independent and scales primarily with the
HCF fraction and route length. This means architectural choice
(1+1 vs.\ SBPP) offers no leverage in reducing the CO$_2$ burden;
mitigation must be addressed at the physical layer through gas
management or spectral pre-equalization~\cite{sillekens_ofc2026}.

\section{Mitigation Strategies and Network Design Guidelines}

We outline a mitigation framework spanning DSP, physical layer, and
network planning.

\subsection{DSP Pre-loading Architecture}

A promising mitigation strategy is DSP coefficient pre-loading.
Commercial coherent transceivers
already support configurable CD acquisition windows: platforms such as
Cisco IOS-XR expose \texttt{cd-min}/\texttt{cd-max} parameters that
constrain the DSP's dispersion search range, reducing blind-acquisition
time when the expected accumulated CD of the path is known in
advance.  The
same mechanism can be extended to protection switching: the network
controller maintains a protection path database containing, for each
potential switchover target, the accumulated CD in ps/nm, the expected
PMD, the IMI level and differential group delay, and the CO$_2$
absorption profile (if the protection path contains HCF). Upon detection of a working-path failure, the receiver DSP
narrows its equalizer search window to the pre-computed range before
training on the live signal, avoiding the full blind CD sweep that
would otherwise be required. 

Fast coherent receiver acquisition has been demonstrated in burst-mode
WDM systems: Thomsen~et~al.\ report burst acquisition times below
200~ns in a back-to-back five-channel system~\cite{Thomsen2011}, achieved through efficient equalizer initialization algorithms rather than coefficient pre-loading per se. While the specific convergence
time for cross-fiber HCF--SMF switching has not yet been
experimentally measured, the qualitative conclusion --- that
pre-loading reduces equalizer adaptation requirements ---
is well established.

For CO$_2$ absorption management, the spectral pre-equalization
approach~\cite{sillekens_ofc2026} can be pre-computed and stored as
a per-channel transmitter-side filter.  When protection is invoked to
an HCF path, the transmitter activates the pre-computed spectral
pre-equalization filter simultaneously with the path switch. This
approach is, however, constrained by the available EDFA output power:
pre-emphasizing the channels affected by CO$_2$ absorption requires
boosting their launch power relative to unaffected channels, which
increases the total per-span power and can drive the EDFA into
saturation, limiting
the degree of pre-equalization that can be applied. 

\subsection{Hitless Switching and Buffer-Aided Protection}
A complementary mitigation to DSP pre-loading is \emph{hitless}
(or \emph{errorless}) switching. In standard 1+1 protection, the
transmitter bridges identical data onto both working and protection
paths simultaneously. A hitless receiver buffers both streams and
aligns them using frame markers or sequence numbers, ensuring
error-free output as long as one uncorrupted copy of each data unit
arrives from either path.

The delay asymmetry challenge identified in
Section~\ref{sec:delay_asymmetry} can be partially converted into a
mitigation opportunity. For the worst-case CORONET pair (280~km
working, $\sim$5,000~km protection), routing the protection path over
HCF rather than SMF reduces the differential delay from $\sim$23~ms
to $\sim$15~ms --- a 34\% reduction that translates directly into
34\% smaller hitless buffers. At 400~Gb/s, this saves $\sim$400~MB
of high-speed memory per protected channel. Additionally, HCF's
thermal coefficient of delay is $\sim$20$\times$ smaller than SMF,
stabilizing the differential delay over time and simplifying buffer
alignment tracking.

From a planning perspective, preferentially deploying HCF on the
longer links of a topology --- which are statistically more likely to
appear in protection paths --- simultaneously improves GSNR and
reduces hitless buffer requirements, providing an additional
criterion for HCF link selection beyond latency and capacity.

\subsection{Design Considerations}


A key consideration is managing the $\sim$1~GHz-wide CO$_2$ 
absorption notches present in HCF L-band transmission~\cite{chen_ecoc2025,he_ofc2026}. 
Two primary mitigation pathways exist:

\textbf{Gas management.}
Post-fabrication gas purging is practical for short lengths but scales 
poorly for telecom distances and is easily undone by non-hermetic field 
splices. A more robust approach is fabrication-level environmental control. 
By excluding CO$_2$ from the draw atmosphere and hermetically sealing the 
cabled fiber with SMF pigtails, absorption is suppressed at the source. 
For example, Microsoft's recently demonstrated Hybrid-DNANF achieved a 
$\sim$4$\times$ reduction in CO$_2$ absorption through fabrication improvements 
alone.

\textbf{Gas-insensitive operation near 1~$\mu$m.}
Because CO$_2$ lacks significant absorption between $\sim$900--1100~nm, 
shifting transmission to the 1~$\mu$m window provides intrinsic immunity. 
The Hybrid-DNANF achieved 0.13~dB/km at 1015~nm and 0.11~dB/km at 1550~nm, 
offering 74~nm of bandwidth below 0.2~dB/km. The supporting 1~$\mu$m 
ecosystem is rapidly maturing, leveraging Yb-doped fiber amplifiers, GaAs 
lasers, and recently demonstrated $>$100~GHz TFLN modulators achieving 
160~Gb/s PAM4. However, practical deployment requires establishing a 
standardized 1~$\mu$m DWDM grid and characterizing IMI in this window, which currently remains unquantified.

\subsection{Network Planning Principles}

Our simulation results yield the following planning guidelines:

\textbf{Principle 1: DSP pre-loading is critical}, for
any network containing both fiber types. This can potentially be a procurement
requirement for coherent transceivers intended for hybrid networks.

\textbf{Principle 2: Prioritize symmetric deployment}.  When choosing
which links to upgrade, prefer strategies that maintain fiber-type
homogeneity along diverse path pairs.  Upgrading both the working and
protection path's links simultaneously avoids creating cross-fiber
protection scenarios.

\textbf{Principle 3: Characterize and store protection path profiles}.
Before commissioning any HCF link, it is essential to measure its full spectral characterization (including CO$_2$ absorption profile, IMI level, and CD coefficient)~\cite{he_ofc2026} and store it in the network management system. This information feeds the DSP pre-loading database.

\textbf{Principle 4: Account for the DSP complexity gap in SLA design}.
Service level agreements should acknowledge that cross-fiber protection
events require substantially larger equalizer adaptation ranges (up to
3,300 additional FDE taps for worst-case node pairs) than same-fiber
events.

\textbf{Principle 5: Plan for full deployment as the end state}.
Our results show that at 100\% HCF deployment, the cross-fiber problem
improves significantly, and capacity retention reaches 85--99\%.

\textbf{Principle 6: Prefer 1+1 over SBPP for hybrid deployments}.
Our comparative analysis demonstrates that SBPP consistently performs
worse than 1+1 in hybrid HCF--SMF networks.  SBPP's greedy
shortest-first path selection creates greater path-length asymmetry
(Fig.~\ref{fig:arch}), leading to higher CD steps and GSNR penalties.
More critically, SBPP can fail entirely on trap topologies---we
identified two such failures in COST~266.  While SBPP improves
spectral efficiency in homogeneous networks through backup capacity
sharing, this advantage is offset by the increased DSP adaptation
burden in hybrid deployments.  Furthermore, SBPP does not pre-provision
the protection path, meaning the receiver cannot pre-monitor the
protection signal; this makes DSP pre-loading (Principle~1) more
challenging.

\section{Open Research Challenges}

Several important questions remain for future investigation.

\textbf{Dynamic gas absorption:} The CO$_2$ absorption profile may
change over the fiber's lifetime due to gas diffusion, temperature
variations, or micro-leak ingress through cable joints.  Water
ingress into the hollow core can occur within minutes of exposure,
making hermetic cable sealing an ongoing operational requirement.
Long-term monitoring and adaptive pre-equalization strategies are
needed, potentially incorporating periodic spectral sweeps to update
the DSP pre-loading database as the absorption landscape evolves.

\textbf{Multi-vendor interoperability:} Current HCF designs vary
significantly across manufacturers in air-core diameter, mode field
diameter, number of cladding tubes, and operating wavelength window.
Standardization of HCF physical parameters through bodies such as
ITU-T SG15 (targeting its first technical report on HCF by July 2026),
IEEE 802.3 (investigating HCF impact on 1.6~TE Ethernet modules), and
CCSA (which has launched five research projects on HCF since 2022) is
essential for protection switching across multi-vendor domains.

\textbf{Experimental validation of cross-fiber switching:} While
commercial coherent DSPs support CD pre-provisioning for wavelength
commissioning, no published experiment has measured the actual DSP
reconvergence behavior when switching between an SMF path and an HCF
path.  The unique combination of simultaneous CD step, IMI onset, and
potential gas absorption transient has not been characterized.
Experimental demonstrations using commercial 400ZR+ or open-line-system
transponders on hybrid testbeds are urgently needed to validate DSP
pre-loading effectiveness for cross-fiber scenarios.

\textbf{Bidirectional transmission on a single HCF strand:}
HCF's ultra-low Rayleigh backscatter enables same-wavelength
bidirectional transmission on a single fiber strand, recently
demonstrated at 400G~ZR over hollow-core cable~\cite{hong_ofc2026}.
In conventional SMF networks, Rayleigh backscatter from the
counter-propagating signal accumulates over long spans and degrades
receiver sensitivity, forcing operators to dedicate separate fiber
strands to each traffic direction. HCF's air-core guidance reduces
backscatter by several orders of magnitude relative to SMF, making
the crosstalk from the counter-propagating signal negligible at the
coherent receiver. This allows a single HCF strand to carry both
traffic directions simultaneously, effectively halving the fiber
count required for a bidirectional link and reducing infrastructure
cost. From a protection perspective, this infrastructure saving can
be reinvested: the strand that would otherwise have been consumed by
the return direction becomes available as a dedicated protection
fiber, potentially enabling geographically diverse protection on
routes where fiber count is the binding constraint. It does not,
however, eliminate the cross-fiber DSP adaptation challenge, since
the protection path may still traverse a different mixture of fiber segments than the working path.

\textbf{Architecture selection and fiber-aware routing:} While this study 
focuses on a comparison between 1+1 protection and SBPP, the architectural 
design space for hybrid SMF-HCF networks is significantly broader. Future 
work should evaluate alternative protection schemes, such as pre-configured 
cycle protection (\textit{p}-cycles) and shared mesh restoration, which may 
yield different trade-offs regarding capacity retention and spectral efficiency 
in mixed-fiber environments. Furthermore, exploring adaptive architectures---where 
the control plane dynamically selects between 1+1 and SBPP based on the specific 
fiber composition of the candidate working and protection paths---could maximize 
the spectral efficiency of SBPP for homogeneous path pairs while falling back 
to 1+1 to mitigate the cross-fiber penalty on mixed paths. Crucially, the 
performance of any protection architecture in a hybrid network is fundamentally 
constrained by the underlying routing algorithms. Developing novel fiber-aware 
and multi-impairment aware routing protocols is essential. These algorithms must 
go beyond simple distance or homogeneous-GSNR metrics to explicitly model the 
transition penalties (e.g., DSP reconvergence time, latency asymmetry, and 
differing nonlinear thresholds) incurred when switching between SMF and HCF segments.

\textbf{Real-time DSP pre-loading protocols:} The signaling mechanism
for transferring protection path characteristics to the receiver DSP
is not yet standardized.  Integration with existing GMPLS/RSVP-TE or
software-defined networking control planes is needed.  The protocol
must support rapid dissemination of updated path parameters whenever
the network topology changes.

\section{Conclusion}

As hollow-core fibers transition from laboratory demonstrations to field deployments, optical network operators face the practical challenge of managing hybrid HCF--SMF infrastructure. This article has provided a comprehensive overview of the protection switching complexities that arise in such mixed-fiber environments. 

We identified the primary physical-layer challenges introduced by cross-fiber switching, including massive, instantaneous steps in accumulated chromatic dispersion, the onset of IMI, L-band gas absorption penalties, and launch power asymmetries. 

Through Monte Carlo simulation across six reference topologies, we quantified these impairments under both 1+1 dedicated protection and Shared Backup Path Protection (SBPP). Our analysis revealed that cross-fiber switching is fundamentally asymmetric: HCF$\rightarrow$SMF transitions inflict severe GSNR penalties and CD steps, while SMF$\rightarrow$HCF transitions can often improve signal quality. Furthermore, our comparison established that 1+1 dedicated protection is significantly better suited for hybrid networks than SBPP. The joint path optimization of 1+1 minimizes path-length asymmetry and DSP adaptation requirements, whereas SBPP's greedy routing approach exacerbates impairments and introduces trap-topology failures.

To manage the interim hybrid deployment phase, we proposed a mitigation framework centered on DSP coefficient pre-loading, hitless buffer-aided protection, and symmetric network planning principles. Ultimately, capacity retention improves monotonically as HCF penetration increases, motivating full deployment as the long-term target.

Finally, we highlighted critical open research areas necessary to fully mature hybrid network operations. These include developing dynamic compensation for gas absorption, experimental validation of cross-fiber DSP reconvergence transients, standardization for multi-vendor interoperability, and the design of novel fiber-aware routing algorithms capable of directly mitigating cross-fiber penalties.
	\section*{Disclaimer and Acknowledgment}
The views and opinions expressed in this document belong solely to the author and do not reflect Huawei's official stance. Generative AI was utilized to refine language and grammar.
\bibliographystyle{IEEEtran}
\bibliography{hcf_protection_network}

\begin{IEEEbiography}{Md Ghulam Saber} received
the bachelor’s and Master of Science degrees in electrical and
electronic engineering from the Islamic University of Technology
(IUT), Bangladesh in 2013 and 2015, respectively, and the Ph.D. degree
from McGill University, Montreal, QC, Canada, in 2019. He
was honored with the IUT Gold Medal for academic excellence
during his bachelor’s studies. He received the R. H. Tomlinson
Doctoral Fellowship and the FRQNT Doctoral Fellowship
from the province of Quebec. In 2019, he also secured the SPIE
Optics and Photonics Education Scholarship and the esteemed
IEEE Photonics Society Graduate Student Scholarship.
\end{IEEEbiography}

\begin{IEEEbiography}{Zhiping Jiang} received the bachelor’s, master’s, and Ph.D.
	degrees from the National University of Defense Technology
	(NUDT), China. He worked on THz generation and detection
	from 1997 to 2000. He was a Senior Designer in optical modem
	technologies at Nortel Networks from 2000 to 2010. He then
	joined Huawei Technologies Canada. He is currently a Distinguished
	Engineer. His research interests include performance
	modeling, optimization, and monitoring in fiber communication
	systems.
\end{IEEEbiography}
\end{document}